\documentclass[10pt,conference]{IEEEtran}
\IEEEoverridecommandlockouts

\usepackage{graphicx}
\usepackage{booktabs}
\usepackage{amsmath,amssymb}
\usepackage{subcaption}
\usepackage{listings}
\usepackage{xcolor}
\usepackage{tikz}
\usetikzlibrary{arrows.meta,positioning}
\usepackage[hidelinks]{hyperref}
\usepackage{cite}
\usepackage{balance}
\usepackage{ifthen}
\usepackage{xspace}

\newboolean{showcomments}
\setboolean{showcomments}{true}         
\ifthenelse{\boolean{showcomments}}
  {\newcommand{\nb}[2]{
  \fbox{\bfseries\sffamily\scriptsize#1}
     {\sf\small$\blacktriangleright$\textit{\textcolor{red}{#2}}$\blacktriangleleft$}
   }
  }
  {\newcommand{\nb}[2]{}
   
  }

\newcommand{\bridge}{PCLA-Bridge\xspace}

\begin{document}

\title{\bridge: A Framework for Cross-Domain Deployment of CARLA Agents}

\author{\IEEEauthorblockN{Masoud Jamshidiyan Tehrani\IEEEauthorrefmark{1},
Jinhan Kim\IEEEauthorrefmark{1}, Paolo Tonella\IEEEauthorrefmark{1}}
\IEEEauthorblockA{\IEEEauthorrefmark{1}Universit\`a della Svizzera italiana, Lugano, Switzerland\\
\{masoud.jamshidiyantehrani, jinhan.kim, paolo.tonella\}@usi.ch}}


\maketitle

\begin{abstract}
Most high-performing end-to-end autonomous driving agents (ADAs) are developed for the CARLA Leaderboard, tightly coupling them to CARLA's APIs, sensor layouts, coordinate systems, and internal clock. Reusing these models on real datasets, alternative simulators, or physical platforms necessitates substantial manual re-engineering. We present \bridge, a deployable tool that runs unmodified CARLA Leaderboard agents from any sensor source to any control destination without an active CARLA server. With \bridge\ handling dependency isolation, tensor marshalling, routing, and synthetic timing, integrators only need to provide a minimal interface. We validate the tool across two operational paradigms: offline replay on in-domain CARLA recordings and real-world nuScenes logs, and live closed-loop control across three diverse simulation engines: MetaDrive, AWSIM, and AlpaSim. We analyze the lightweight integration requirements for each platform, highlighting output translation (such as converting agent actions into target trajectory formats) and coordinate-frame transformations, showing that integration requires minimal glue code rather than structural changes. All eight evaluated agents execute across all three simulators without modification, confirming the reliability of \bridge\ and establishing a foundation for systematic sim-to-real, cross-sim, and security testing.
\end{abstract}


\section{Introduction}
End-to-end autonomous driving agents (ADAs) are typically built and evaluated in a single simulator, most commonly the CARLA simulator~\cite{carla}. Because these implementations assume a specific end-to-end environment, relying on native sensor tensors, routing, temporal synchronization, and coordinate conventions, a published agent becomes tightly coupled to its original execution harness. This coupling hinders code reuse, cross-environment evaluation, and any transition toward physical deployment. Currently, to run a published CARLA agent on real-world logs or within another simulator, researchers must manually re-engineer the agent.

While the PCLA framework~\cite{pcla} decoupled 41 Leaderboard agents from the Leaderboard codebase~\cite{leaderboard} to enable execution across diverse CARLA environments, those agents remained bound to the simulator itself. This paper offers a tool that removes that dependency. \bridge\ executes unmodified agents without a CARLA server, allowing any data source to drive the agent and any control target to consume its outputs. While an earlier extended abstract introduced this concept using offline replay~\cite{pclabridge-src}, this paper describes a fully functional tool featuring live, closed-loop operation and documents the practical requirements for integrating new data sources.

\noindent The envisioned users are autonomous driving (AD) researchers and software engineering (SE) testers who need to evaluate a published PCLA agent on inputs that the agent's native simulator cannot produce. These inputs may come from real-world logs, alternative simulators, adversarial perturbations, or physical vehicles. \bridge\ enables this evaluation without requiring the user to modify or rewrite the agent.


\section{\bridge}
\label{sec:tool}
The insight behind \bridge is that an agent's per-frame execution is inherently simulator-agnostic:

\begin{lstlisting}
control = agent.run_step(sensor_dict, timestamp)
\end{lstlisting}

\noindent where \texttt{sensor\_dict} maps sensor identifiers to data arrays, and \texttt{control} is a tuple containing throttle, steer, and brake values. The vendored \texttt{carla} module is required solely to resolve data types rather than to maintain an active server connection.

\begin{figure}[t]
  \centering
  \includegraphics[width=\linewidth]{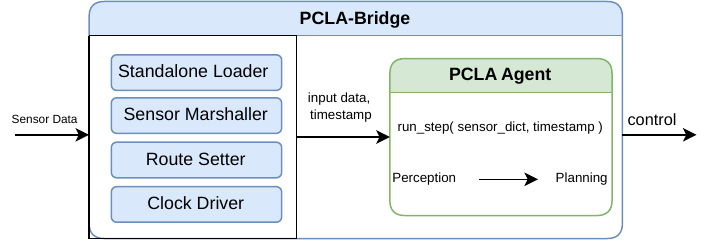}
  \caption{PCLA-Bridge architecture.}
  \label{fig:arch}
\end{figure}

As illustrated in Fig.~\ref{fig:arch}, \bridge\ loads each agent in a \textbf{standalone} environment. This isolates the agent's specific dependencies and ensures that mutually incompatible library versions never collide. The bridge then handles the complex data transformations that an integrator would otherwise need to implement manually. Specifically, it \textbf{marshals} external sensor readings into the exact tensor layout expected by the agent by resizing cameras and zero-filling absent modalities. It also \textbf{sets} the navigation route using GPS waypoints and manages a \textbf{synthetic clock} for agents that rely on wall time. Because both the data source (such as a simulator, a dataset log, or a physical vehicle) and the control target are entirely external, \bridge\ is inherently source- and target-agnostic.

As detailed in Table~\ref{tab:tax}, portability is fundamentally a property of the agent rather than the bridging tool. Sensor-only agents consume physical data streams and can be ported directly to new environments. Privileged agents, however, rely on ground truth data that only the host simulator can provide, such as exact ego pose or rendered bird's-eye view, meaning they cannot be driven by empirical sensors. Finally, a third class of systems, such as Autoware~\cite{autoware}, does not follow a monolithic sensor-in and control-out architecture and therefore cannot be ported using this approach.

\begin{table}[t]
\caption{Reusability taxonomy of the PCLA agent library.}
\label{tab:tax}
\centering\footnotesize
\begin{tabular}{@{}p{2.4cm}p{5.2cm}@{}}
\toprule
\textbf{Class} & \textbf{Agents (representative)}\\
\midrule
Sensor-only (portable) & TransFuser v3--v6, InterFuser, NEAT, LAV, WOR, LBC, ThinkTwice,
SimLingo, LMDrive, ORION, MindDrive \cite{transfuser,hiddenbiases,tfv6,interfuser,neat,lav,wor,lbc,thinktwice,simlingo}\\
Privileged (not portable) & PlanT, PlanT2, CaRL, Roach \cite{plant2,carl}\\
Not a \texttt{run\_step} model & Autoware (external ROS\,2 stack) \cite{autoware}\\
\bottomrule
\end{tabular}
\end{table}

\section{Related Work}

The most relevant works from the literature deal with the sim-to-real gap and with the limitations of existing ADAs benchmarks.
Several works from the literature attempt to close the sim-to-real gap via generative perception bridging~\cite{sim2realdiff}, 
Gaussian splatting for high-fidelity sensor rendering~\cite{nurec}, 
or policy adaptation~\cite{sim2realad}. However, recent work shows that agents achieving top CARLA benchmark scores still transfer poorly to reality~\cite{carlarlreview}.
%
%
Moreover, industrial surveys rank sim-to-real discrepancies and sensor fidelity among the most critical testing challenges~\cite{adsreality}. Reality-gap analyses further confirm that discrepancies span perception, actuation, and vehicle behavior. Perception-focused test inputs frequently miss system-level failures~\cite{realitygapmm}. 

Open-loop benchmarks like nuScenes frequently suffer from data leakage and are biased toward simple straight paths, allowing simple near-constant velocity baselines to outperform more sophisticated learned models~\cite{egostatus}. NAVSIM evaluates rule-based trajectories on recorded data as a pragmatic compromise due to the overhead of closed-loop testing~\cite{navsim}. Both frameworks only evaluate predicted trajectories on static benchmarks. While PCLA~\cite{pcla} improved agent testability, it remained constrained to CARLA.


\section{Bridging to a new simulation}
\label{sec:integrate}

We evaluate on three simulators chosen to span distinct autonomous driving paradigms. (\textbf{1) MetaDrive}~\cite{metadrive} is a lightweight, Python-native engine (Panda3D) featuring procedurally generated, non-photorealistic scenes and depth-approximated pseudo-LiDAR. \textbf{(2) AWSIM}~\cite{awsim} is a photorealistic Unity digital twin integrated with the production Autoware and ROS 2 stacks, providing genuine GPU-accelerated LiDAR. \textbf{(3) AlpaSim}~\cite{alpasim} utilizes NVIDIA NuRec 3D Gaussian splatting~\cite{nurec} 
to render camera-only neural reconstructions directly from physical driving logs. These simulators vary across every critical dimension: rendering style (procedural to game-engine to neural), scene origin (synthetic to digital twin to real-world logs), software stack (Python API to Unity/ROS 2 to gRPC microservices), and sensing modalities. In order of increasing perceptual realism, they form a structured progression from CARLA toward physical deployment, closing the gap between offline replays and real-world execution.

Instantiating \bridge for a new simulation source requires the creation of a small adapter. With the automated capabilities that \bridge already provides, the integrator needs to implement only three lightweight components: control-output translation, frame calibration, and sensor mapping. In our integration, enabling all three simulators required precisely these three adjustments and no further modifications.

\textbf{Control-output translation.} While an agent outputs a standard tuple of throttle, steer, and brake, target simulators often require distinct control formats. CARLA and MetaDrive consume this triplet directly. In contrast, AWSIM's Autoware interface expects a steering-tire angle alongside pedal actuations, and AlpaSim's external driver interface requires a planned trajectory instead of instantaneous control inputs. For Alpasim, a lightweight adapter scales the agent's steering input ($\in[-1,1]$) to a wheel angle using the vehicle's maximum steering lock, while translating throttle and brake commands into a target speed. It then integrates a kinematic motion model forward over a short horizon to produce a constant-curvature arc of waypoints. Because this arc is recomputed at every cycle, it serves strictly to convey the agent's instantaneous intent, leaving low-level path tracking to AlpaSim's internal controller. In this way, \bridge accommodates three structurally distinct simulators with only a few lines of translation logic for each.

\textbf{Frame calibration.} Each simulation engine maintains its own coordinate and heading conventions. The frame adapter reconciles these differences using a compact set of constants, such as a GPS $y$-axis inversion, compass orientation offsets, and steering sign conventions, each calibrated once against a known-good reference drive. 
For instance, an initial sign discrepancy in AlpaSim's ego-pose handling component created inconsistencies with the synthetic GPS signal supplied to the agent. Once corrected, the two signals became consistent with each other.

\textbf{Sensor mapping.} A target data source does not need to expose every modality declared by an agent. For example, AlpaSim's driver interface provides camera feeds and egomotion estimates but omits LiDAR. In response, \bridge\ automatically zero-fills the missing LiDAR channels, allowing multimodal camera-LiDAR agents to gracefully degrade and operate in a camera-only configuration.

Because both the data source and the simulation environment remain completely external, one advantage offered by \bridge is that the ADA's agent can execute within its own process, even on an isolated host. In our AlpaSim integration, we ran the agent on one machine and the simulator's GPU-accelerated renderer on another, bridging communication over a network socket routed through an SSH tunnel. The source-agnostic architecture supported this distributed deployment without requiring any structural changes to the codebase.

\section{Usage Demonstration}
\label{sec:validate}
We demonstrate the utility of the tool across two operational settings: offline, to assess whether agents can navigate from arbitrary recorded data streams, and live, to evaluate closed-loop execution within foreign simulation environments. For our empirical evaluation, we select eight sensor-only agents spanning diverse model architectures with two input modalities. Four are camera-only TransFuser v4 (aim), LBC, NEAT, and the SimLingo vision-language-action model, and four are camera+LiDAR TransFuser v4 (lav), TransFuser v5, ThinkTwice, and InterFuser. Because all eight are top-tier Leaderboard submissions, observed performance differences isolate agent portability rather than underlying deficiencies.

\begin{figure*}[t]
  \centering
  \begin{subfigure}[b]{0.48\linewidth}
    \centering
    \includegraphics[width=\linewidth, trim={0 112pt 0 0pt}, clip]{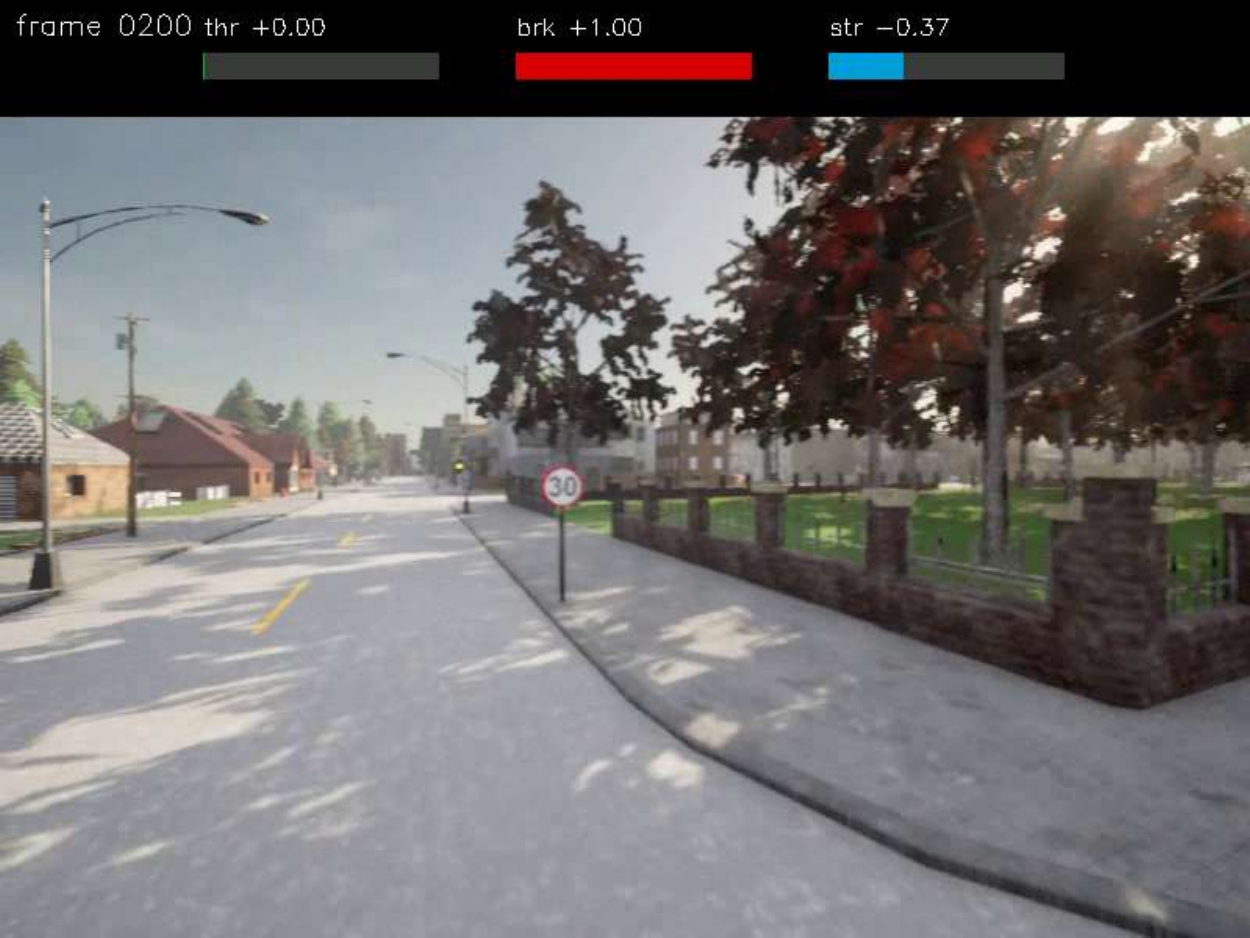}
    \caption{CARLA replay}
    \label{fig:carla}
  \end{subfigure}
  \hfill
  \begin{subfigure}[b]{0.48\linewidth}
    \centering
    \includegraphics[width=\linewidth]{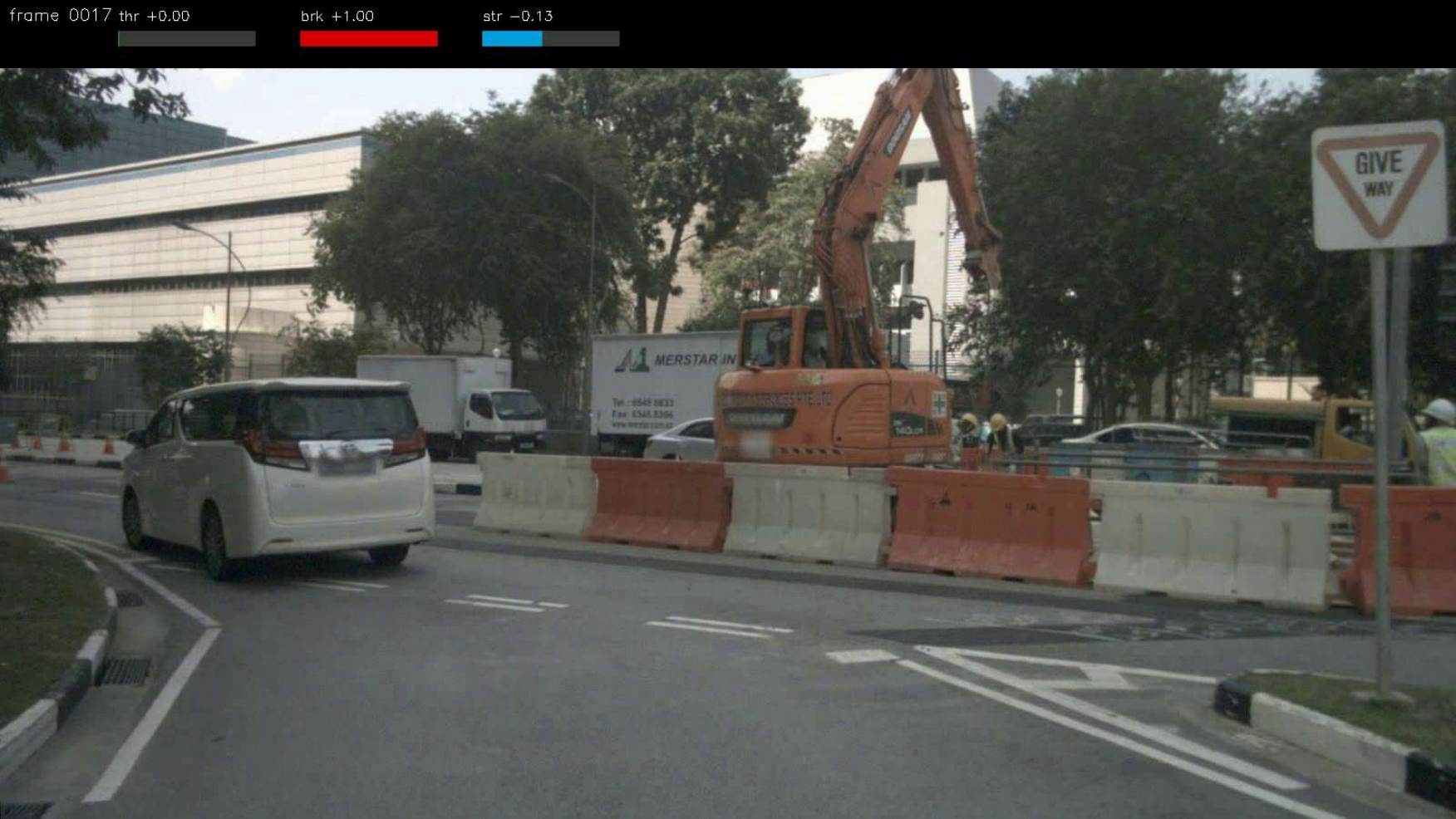}
    \caption{nuScenes replay}
    \label{fig:nuscenes}
  \end{subfigure}
  \caption{Bridge evaluation across synthetic and real-world drives.}
  \label{fig:offline}
\end{figure*}

\textbf{Offline replay (open-loop).} When replaying recorded drives through the bridge and inspecting the resulting control commands, every agent could load and execute without requiring a CARLA server. As illustrated in Fig.~\ref{fig:offline}, the generated control accurately tracks the planned route on an in-domain CARLA recording. When tested on real-world data from the nuScenes dataset~\cite{nuscenes}, where real front-camera frames and ego poses are converted to the bridge's input format, the CARLA-trained agent correctly steers left through the scene's actual left turn. At the same time, the sim-to-real gap manifests itself as near-zero throttle on unfamiliar real-world imagery, reflecting an overly cautious policy. Overall, this demonstrates that the tool ingests empirical sensor logs and produces coherent, domain-responsive control commands.

\textbf{Live sim-to-sim (closed-loop).} We then evaluated the agents live, applying the returned control commands at every time step across the three selected simulation platforms.
For each simulator, we construct both a baseline scene A and a more challenging scene B. Every portable agent executes unmodified across all three engines using solely the lightweight adapter described in \S\ref{sec:integrate}. 
Table~\ref{tab:results} summarizes the route coverage metrics. The fact that these agents operate successfully across fundamentally distinct software stacks validates the tool's portability. 


\begin{table}[t]
\caption{Route coverage (on-route distance, m) per agent across three simulators and two live scenarios}
\label{tab:results}
\centering\footnotesize\setlength{\tabcolsep}{4pt}
\begin{tabular}{@{}l cc cc cc@{}}
\toprule
& \multicolumn{2}{c}{\textbf{MetaDrive}} & \multicolumn{2}{c}{\textbf{AWSIM}} & \multicolumn{2}{c}{\textbf{AlpaSim}}\\
\cmidrule(lr){2-3}\cmidrule(lr){4-5}\cmidrule(lr){6-7}
\textbf{agent} & A & B & A & B & A & B\\
\midrule
\texttt{tfv4\_lav} & 195 & 197 & 40 & 39 & 80 & 40\\
\texttt{tfv4\_aim} & 123 & 43 & 22 & 30 & 80 & 39\\
\texttt{tt}        & 122 & 156 & 30 & 107 & 66 & 72\\
\texttt{if}        & 122 & 28 & 7 & 86 & 40 & 39\\
\texttt{tfv5}      & 116 & 117 & 163 & 24 & 13 & 42\\
\texttt{simlingo}  & 130 & 73 & 0 & 11 & 80 & 39\\
\texttt{lbc}       & 85 & 85 & 18 & 0 & 39 & 40\\
\texttt{neat}      & 42 & 18 & 0 & 9 & 26 & 38\\
\bottomrule
\end{tabular}
\end{table}

\section{What \bridge\ Enables}
\label{sec:enables}
With CARLA agents directly executable on arbitrary input sources, multiple software engineering and autonomous driving evaluations shift from re-engineering tasks to one-line integrations:
\begin{itemize}
    \item \textbf{Quantifying the sim-to-real gap:} Executing CARLA-trained policies on real-world datasets, such as nuScenes or Waymo, directly measures behavioral degradation, as evidenced by the near-zero throttle hesitation observed under empirical camera imagery.
    \item \textbf{Security and adversarial transferability:} Feeding physical attack sensor streams into the same models determines whether vulnerabilities discovered inside CARLA persist under real-world conditions.
    \item \textbf{Cross-simulator fault localization:} Keeping the agent architecture fixed while swapping the underlying simulation engine isolates whether a driving violation stems from model defects or from host-specific physics and rendering anomalies.
    \item \textbf{Scalable offline regression testing:} Replaying thousands of kilometers of recorded sensor trajectories verifies behavioral consistency without the steep compute overhead of a GPU-bound Unreal Engine instance.
    \item \textbf{Benchmark standardization:} Evaluating diverse, heterogeneous models under unified execution standards establishes a consistent testing ground.
\end{itemize}

The scene-dependent variance in Table~\ref{tab:results}, where no single agent dominates across all engines, highlights the empirical diagnostic signals this framework can systematically measure. 

\section{Conclusion and Future Work}
\bridge\ is a deployable tool that makes CARLA-trained autonomous driving agents reusable without an active CARLA simulation in the loop, focusing on sensor-only agents from the PCLA framework. To demonstrate its practical utility, we showed that the tool reliably generates control outputs from both offline recorded data and external simulation platforms. Specifically, we evaluated the framework on recorded in-domain CARLA executions as well as real-world nuScenes sequences. Furthermore, we deployed unmodified agents in closed-loop control across three distinct foreign simulators, demonstrating basic driving capabilities despite domain-shift performance variations. 

In future work, we plan to validate the tool on a physical vehicle, formulate behavioral fidelity and portability metrics to systematically determine why certain architectures transfer better outside CARLA, and conduct adversarial transferability studies across sim-to-sim and sim-to-real boundaries to uncover intrinsic agent vulnerabilities without simulator biases.

\section*{Data Availability}
The tool is available at \url{https://github.com/MasoudJTehrani/PCLA/tree/main/pcla_bridge}. Validation artifacts are available at \url{https://huggingface.co/datasets/MasoudJTehrani/pcla-bridge-evaluations}. The demonstration video is at \url{https://youtu.be/Z0uGPAGPR8c}.

\bibliographystyle{IEEEtran}
\bibliography{references}

\end{document}